# THEORETICAL SETTING OF INNER REVERSIBLE QUANTUM MEASUREMENTS

Paola A. Zizzi

Dipartimento di Matematica Pura ed Applicata
Università di Padova
Via Belzoni, 7
35131 Padova, Italy
zizzi@math.unipd.it

**Abstract**

 We show that any unitary transformation performed on the quantum state of a closed quantum system describes an inner, reversible, generalized quantum measurement. We also show that under some specific conditions it is possible to perform a unitary transformation on the state of the closed quantum system by means of a collection of generalized measurement operators. In particular, given a complete set of orthogonal projectors, it is possible to implement a reversible quantum measurement that preserves the probabilities.
In this context, we introduce the concept of "Truth-Observable", which is the physical counterpart of an inner logical truth.

# 1. Introduction

The original measurement postulate of quantum mechanics, introduced by von Neumann [1], considered only projective measurements, where a "standard" quantum observable *A* has a spectral resolution in terms of orthogonal projection operators. The postulate states that during a measurement of *A*, the state vector of the quantum system reduces to the eigenvector of *A* corresponding to the measurement result. More recently, the notion of quantum measurement has been extended to generalized measurements [2] for which the concept of "generalized" quantum observable has been introduced, where the positive resolution of the identity replaces the spectral decomposition. The concept of a generalized quantum measurement is necessary to describe the quantum interaction between a quantum system and a measuring device. Moreover, the generalized measurement formalism can deal with phenomena not captured by projective measurements; in particular, it is quite useful for describing joint measurements of incompatible "standard" observables.

Finally, a generalized formalism appears to be more suitable for quantum computing (for a review of quantum computing see ref. [2]).

However, although generalized measurements can be implemented by the use of projectors augmented with unitary evolution, and ancillary system, one does not really overcome the measurement problem.

In the case of quantum computers, the above problem becomes even worse, as the irreversibility of a quantum measurement is dramatically incompatible with the reversibility of the whole computational process.

Thus, one should find a way to bind quantum measurement and quantum evolution/computation in a single mathematical formulation.

We believe that this unification is possible only when the "observer" is conceived internal to the original quantum system to be measured. To do so, the "observer" should enter a quantum space background whose quantum states are in one-to-one correspondence with the computational states of the quantum computer [3].

In some sense, this idea is the analogous of Feynman's idea [4] that the most efficient way to simulate a quantum system is to use another quantum system (a quantum computer). In our case, the most adequate way to simulate a reversible measurement is to use another quantum system, which is not the apparatus, but the quantum space background of the quantum system itself, as illustrated in [3].

The results of [3] have also been exploited in quantum gravity [5] and in quantum computability [6].

The present paper is conceived manly to support the ideas and results of [3] by formalizing the notion of internal reversible measurement in the framework of the generalized measurement formalism. As our inner reversible quantum measurement is nothing but a unitary transformation, here we show that any unitary transformation is formally a generalized quantum measurement.

As mentioned above, the theory of generalized measurement has been developed mainly to cope with some peculiar features of quantum computing. However, the so-called quantum computational logics [7] adopt (following the traditional quantum logic) the old projective measurement scheme. In this way, in our opinion, one can achieve only a partial logical structure of quantum computation. On the other hand, in a recent paper [8] we introduced a new kind of quantum logic: the *inner quantum logic* based on the reversible, inner quantum measurement described in [3], which is not projective. The main features of the inner quantum logic are paraconsistency [9] and symmetry, which are also present in basic logic [10].

The question was whether this new kind of non projective measurement could be recognized as a generalized measurement. In the affirmative, the inner quantum logic would relay on proper foundations of quantum mechanics (and quantum information). This is the aim of the present paper. Moreover, we believe that future developments of the inner quantum logic would be relevant also for practical purposes, like program verification of proposed quantum algorithms.

In this paper, we also prove that, once it is given a collection of generalized measurement operators, which satisfy some particular requirements, it is possible to implement a reversible measurement. Such a result might be quite useful in quantum computing, meaning that, at least in principle, one might be able to concatenate quantum measurements in such a way to preserve the reversibility of the whole computational process.

Finally, we introduce the new concept of Truth-Observable, which exhibits a dual nature between physics and logic.

This paper is organized as follows:

In Sect.2, we briefly review Generalized Quantum Measurements (GM), projective measurements (PM) and POVM (Positive Operator-Valued Measure).

In Sect.3, we formally define Inner Reversible Quantum Measurements (IRM) as a special case of GM.

In Sect.4, we discuss the Mirror Measurement (MM) that is a particular case of IRM, which preserves the probabilities.

In Sect.5, we introduce the new concept of Truth-Observable, and its logical meaning.

Sect.6 is devoted to the conclusions.

## 2. Generalized Quantum Measurements (GM): A short Review.

### 2.1 Postulate of GM

A generalized quantum measurement (of a closed quantum system S in the finite dimensional space of states) is described by a collection $\{M_m\}$ of measurement operators on the Hilbert space $H$ of S, where the index $m$ refers to the possible outcomes of the measurement. The measurement operators satisfy the completeness relation:

$$\sum_m M_m^+ M_m = I \qquad (1)$$

Where: $M^+$ is the adjoint of M and $I$ is the identity operator on $H$.

If the state of the system S is $|\Psi\rangle$ just before the measurement, then the probability of getting the outcome $m$ is $p(m) = \langle \Psi | M_m^+ M_m | \Psi \rangle$.

After the measurement, the system S is left in the normalized state: $|\Psi'\rangle = \dfrac{M_m |\Psi\rangle}{\sqrt{p(m)}}$.

The completeness relation (1) ensures that the probabilities sum up to 1: $\sum_m p(m) = 1$ .

### 2.2 Projective measurement (PM) as a particular case of GM

Let us now consider a projective measurement of a "standard" quantum observable $A$ of the system S, which corresponds to a hermitian operator $A = A^+$ on the Hilbert space H of the system S.

The observable has spectral decomposition:

$$A = \sum_m \lambda_m P_m \qquad (2)$$

Where $P_m$ is the projector onto the eigenspace of A with eigenvalue $\lambda_m$. The set $\{P_m\}$ is a complete set of orthogonal projection operators with respect to the observable A, that is, the projectors $P_m$ are hermitian operators: $P_m = P_m^+$, which are orthogonal and idempotent: $P_m P_{m'} = \delta_{mm'} P_m$ and satisfy the completeness relation:

$$\sum_m P_m = I \qquad (3)$$

The possible outcomes of a projective measurement correspond to the eigenvalue $\lambda_m$ of the observable. Upon measuring the system in state $|\Psi\rangle$, the probability of getting the outcome m is $p(m) = \langle \Psi | P_m | \Psi \rangle$.

After the measurement, the system is left in the normalized state: $|\Psi'\rangle = \dfrac{P_m |\Psi\rangle}{\sqrt{p(m)}}$.

A projective measurement is a special case of generalized measurement.
As we have seen in 2.1, the (generalized) measurement operators $M_m$ satisfy the completeness relation (1). If the $M_m$ also satisfy the additional conditions of hermiticity, orthogonality and idempotence, they reduce to projection operators. The postulate of generalized measurement specifies both the probabilities of the different possible outcomes of the measurement, and the post-measurement state. In some cases, however, there is no necessity for knowing the post-measurement state, and interest focuses mainly on measurement statistics. Those cases are best analyzed by using the Positive Operator-Valued Measure (POVM).

**2.3 Definition of POVM**

A POVM operator is defined as:
$$E_m \equiv M_m^+ M_m \qquad (4)$$
The POVM operators satisfy the following properties:
i) The $E_m$ form the positive-definite partition of unity
$$\sum_m E_m = \sum_m M_m^+ M_m = I \qquad (5)$$
ii) The set $\{E_m\}$ is sufficient to determine the probabilities of the different measurement outcomes: $p(m) = \langle \Psi | E_m | \Psi \rangle \equiv tr(E_m \rho)$

Where $\rho = |\Psi\rangle\langle\Psi|$ is the density matrix of the quantum system under consideration.

iii) The $E_m$, are positive operators (hence in particular hermitian): $\langle \Psi | E_m | \Psi \rangle \geq 0$

The $E_m$ as defined in (4) are the "generalized" quantum observables.

Notice that the spectral decomposition (2) of a "standard" observable is replaced, in the generalized case, by the resolution of the identity operator $I$ in (5).
From i) it follows that the probabilities sum up to one, and from iii) it follows that the probabilities are positive:
In case that the measurement operators $M_m$ reduce to projectors $P_m$, then all the POVM elements $E_m$ coincide with the measurement operators $E_m \equiv P_m$.

**3. Inner Reversible Quantum Measurements (IRM)**

**Postulates**

i)   The observer is considered internal to the closed quantum system if and only if the space-time background is a quantum space isomorphic to the quantum system.

ii)   The internal quantum measurement is described by a unitary operator U.

From Postulate i) follows the lemma:

**Lemma 3.1**
The quantum system S and the internal observer O do not form a composite system. Their Hilbert spaces, which are isomorphic, $H_S \approx H_O$ are identified: $H_S \equiv H_O$ and do not form a tensor product, as it is:
$H_{S,O} \equiv H_S \equiv H_O$.

From postulate ii), we get the two following lemma:

**Lemma 3.2**
An internal quantum measurement is a reversible quantum operation.
In fact, any unitary operator U on a Hilbert space $H$ performs a reversible transformation on a quantum state $|\Psi\rangle$ of $H$.

**Lemma 3.3**
In the case the closed quantum system is a quantum computer, the unitary operator U is a quantum logic gate, then an internal quantum measurement coincides with a quantum computational step.

**Theorem 3.4**
A reversible transformation performed by a unitary operator U, on a closed quantum system, which is in the quantum state $|\Psi\rangle$, is a generalized quantum measurement.
Proof:
By definition 3.2, a unitary operator U satisfies the completeness equation (1), in the particular case that the collection $\{M_m\}$ of generalized measurement operators consists of a single operator $M = M_1 \equiv U$.
In this case, the unique outcome m=1 has probability one to occur:
$p_{(1)} = \langle\Psi|U^+U|\Psi\rangle = \langle\Psi|\Psi\rangle = 1$.

After the measurement, the state of the system is: $|\Psi'\rangle = \dfrac{U|\Psi\rangle}{\sqrt{p_{(1)}}} = U|\Psi\rangle$

Then, all the properties of a generalized quantum measurement required by the postulate 2.1 are fulfilled by a unitary operator U.

**Theorem 3.5**
Given a collection $\{M_m\}$ (m = 0, 1, …N-1) of measurement operators, which satisfy the additional condition:
$M_i^+ M_j = M_i M_j^+ = 0$ $(i \neq j), 0 \leq i, j < N$ and a corresponding collection: $\{\alpha_m\}$ of complex numbers satisfying the condition: $|\alpha_m|^2 = 1$, the linear superposition:

$$M = \sum_{m=0}^{N-1} \alpha_m M_m \tag{6}$$

is a unitary operator and then describes a reversible quantum measurement.

The proof is straightforward. We must show that M in (6) is unitary. As M is finite, and it is $\det|M|^2 = 1$, it follows that M is invertible, as $\det M \neq 0$.

**Corollary 3.6**
Given a complete set of orthogonal projection operators $\{P_m\}$ and a corresponding collection of complex numbers $\{\alpha_m\}$ such that $|\alpha_m|^2 = 1$, the linear superposition:

$$\widehat{P} = \sum_{m=0}^{N-1} \alpha_m P_m \tag{7}$$

is a unitary operator (which is a diagonal matrix with respect to the basis of the eigenvectors of the observable A).
$\widehat{P}$ in (7) is the exponential operator: $\widehat{P} = e^{iA}$, where A is the observable, and the complex coefficients $\alpha_m$ in (7) are given in terms of the eigenvalues of A: $\alpha_m = e^{i\lambda_m}$
Then, by theorem 3.4, $\widehat{P}$ describes an inner, reversible, generalized quantum measurement.
Since the projectors $\{P_m\}$ are a special case of $\{M_m\}$, the proof is the same of theorem 3.5.

**Corollary 3.7**
In the case of an IRM, the POVM consists of a single operator. From the definition of $E_m$ it follows:

$$E_1 = M_1^+ M_1 = U^+ U = I \tag{8}$$

and the associated probability is one.

### 4. The Mirror Measurement (MM)

**Definition 4.1**
A mirror measurement $\widehat{M}$ is a unitary transformation on a quantum state $|\Psi\rangle$, which preserves the probabilities of $|\Psi\rangle$.

**4.2 Theorem**
A unitary transformation U on a vector state $|\Psi\rangle$ of a Hilbert space $H$ leaves unchanged the probabilities of $|\Psi\rangle$, if it commutes with a complete set of orthogonal projectors $\{P_m\}$ in $H$.

Proof:
Let us suppose we perform first a unitary transformation U on $|\Psi\rangle$, and then a projective measurement on $|\Psi'\rangle = U|\Psi\rangle$, that is: $P_m|\Psi'\rangle = P_m U|\Psi\rangle$.
The probability of getting the outcome m is:

$p(m)' = \langle \Psi' | P_m | \Psi' \rangle = \langle \Psi | U^+ P_m U | \Psi \rangle$

If it holds: $[U, P_m] = 0$, we get: $p(m)' = \langle \Psi | U^+ U P_m | \Psi \rangle = \langle \Psi | P_m | \Psi \rangle = p(m)$.

**4.3 Corollary**
Any unitary operator on a Hilbert space *H*, which is of the kind $\hat{P}$ given in , preserves the probabilities of a quantum state $|\Psi\rangle$ of *H* on which it acts.

The proof is trivial, as $\hat{P}$ in (6) commutes with the projector operators $P_m$.
Then, by theorem 4.2, when $\hat{P}$ acts on a quantum state $|\Psi\rangle$, it leaves unchanged the probabilities.

**Lemma 4.4**
By definition 4.1 and from corollary 4.2, it follows that every unitary transformation of the kind $\hat{P}$ in (6), is a mirror measurement $\hat{M}$.
The example of mirroring one qubit was given in [3] and [8].

**Theorem 4.6**
Any unitary diagonal matrix of the kind $U_D = e^{i\theta}(\alpha P_0 + \alpha^* P_1)$ on $C^2$ (or its extension to a higher dimensional Hilbert space) describes a GM, and in particular is a mirror measurement $\hat{M}$.

Proof**:**
Any mirror matrix $U_D = e^{i\theta}(\alpha P_0 + \alpha^* P_1)$ is an operator $\hat{P}$ defined in (6),
with: $\alpha_0 = e^{i\theta}\alpha$, $\alpha_1 = e^{i\theta}\alpha^*$, $|\alpha_0|^2 = |\alpha_1|^2 = 1$.
Then, by corollary 3.8, it follows that $U_D$ is a particular case of generalized quantum measurement.
Also, from lemma 4.4, it follows that $U_D$ is a mirror measurement.

The example of mirroring a two qubits state (also, in particular for a Bell state) was given in [11]. If one compares the standard external measurement of a Bell state [2], with our results in [11], one finds the following.
Algebraically, the difference between the internal and the external points of view (concerning a Bell state), stands in the fact that, while in the latter we have two generalized observables, which sum up to the identity:
a) $E_0 + E_1 = I^{(4)}$
In the former, there is only one generalized observable:
b) $E_1 = \hat{M}^+ \hat{M} = I^{(4)}$
From a) we can get only a partial knowledge (information) about the Bell state, while from b), we obtain the whole information at once. In this case, we will say that the single generalized observable is the Truth-Observable that is the topic of next section.

**5. The Truth-Observable**
In mathematics, the word "truth" means "logical truth". On the other side, the word "observable" is a physical concept. But in an inner quantum world, the "truth" is a logical concept, as well as a physical observable, as it is the truth one "observes" inside the closed quantum system.

Let us consider the spectral decomposition (2) in terms of projectors $P_m$ of the "standard" observable A. The observable A reduces to the identity operator $I$ in the n-dimensional Hilbert space $H$ when all the eigenvalues of A are equal to 1, that is the eigenvalue $\lambda = 1$ is n-degenerate. In this case (2) is identified with (3), which is the completeness relation for projectors $P_m$.

More generally, the POVM $\{E_m\}$ realise the positive-definite partition of unity given in (5). The POVM elements $E_m$ are hermitian operators, and are generalized observables (they do not satisfy the orthogonal relation differently from projectors). The logical meaning of both the completeness relation (3) and the positive-definite partition of unity (5) is that the logical truth splits into partial truths, each of them corresponding to an act of measurement from outside.

This is due to the physical fact that any external quantum measurement is an irreversible process, which destroys quantum superposition. Then, an external observer can grasp only fragments of an inner, global truth.

Only an internal observer would be able to achieve the global truth at once, as a whole, by making an IRM, as inside the closed quantum system, he can perform only reversible transformations, described by unitary operators U.

The uniqueness and unitary of such measurement operators allow defining a unique quantum observable that is just the identity.

In fact, in the case that the family of generalized measurement operators $\{M_m\}$ consists of only one element $M_1$ the latter is a unitary operator U and the unique associated generalized observable $E_1$ is the identity operator $I^{(n)}$ in the n-dimensional Hilbert space:

$$E_1 = M_1^+ M_1 = U^+ U = I^{(n)}. \tag{9}$$

Once U is given the meaning of a generalized measurement operator, the identity operator acquires the status of generalized observable, which we will call Truth-Observable.

The Truth-Observable is measured when a quantum operation U is performed on a state $|\Psi\rangle$ of the system followed by the inverse operation:

$$|\Psi\rangle \xrightarrow{U} |\Psi\rangle' \xrightarrow{U^{-1}} |\Psi\rangle \tag{10}$$

In particular, if $|\Psi\rangle$ is a N-qubits state, (10) means that $|\Psi\rangle$ has been computed by a quantum logic gate U, which is a $n \times n$ unitary matrix (where $n = 2^N$) and then un-computed by $U^{-1}$.

In (3) and (5), the truth as a whole can be affirmed only in principle, as the external observer cannot perform all the $\{P_m\}$ (or $\{E_m\}$) operations at the same time. Instead in (9) the truth can be affirmed at once as a whole, as the Truth-Observable is measured in a single step ($E_1$) by the internal observer.

The Truth-Observable is a new concept of a very weak logic, namely paraconsistent, symmetric logic, which turns to be the inner logic of a closed quantum system, like a quantum computer.

Paraconsistency implies (in addition to the invalidation of the excluded middle principle as in intuitionist logic) also the invalidation of the non contradiction principle.

As already mentioned, inner quantum logic is quite alike basic logic. However, differently from basic logic, it has not the (usual) cut rule. In fact, the correct cut rule for our logic seems to be a kind of "branched cut rule" [12]. In physical terms, this means that a single projector is not allowed (but a superposition of projectors is). Thus, by this logic, superposition of states can never be destroyed.
Then, we are looking [12] for a logical calculus whose rules are used in *parallel* (logical calculus in 'branches'). This would focus on the physics of quantum computation in logical terms. In this context, the Truth-Observable is the logical truth based on quantum parallelism, which leads to completeness plus (classical) inconsistency [6].

## 6. Conclusions

We wish to conclude with the following remarks:
i) As we have seen, the inner reversible quantum measurement (IRM) finds its natural setting in the generalized formalism. However, IRM is conceptually very different from all other kinds of measurements considered since now in the literature (GM, PM, POVM). In fact, IRM being internal (while all the others are external) leads to the concept of internal observer (a useful tool for a quantum universe). Also, being reversible (while all the other are irreversible) is consistent, (in the case of a quantum computer) with the computational state.
ii) It should be noticed that, while a GM deals with a composite system (original system plus apparatus) in a higher dimensional Hilbert space, an IRM only deals with a single isolated quantum system.
iii) In the case of the quantum computer, at least, we formally demonstrated the validity of Landauer's principle [13] stating that computation and measurement are fundamentally the same process. In fact, we showed that a quantum logic gate is a special case of (generalized) measurement operator. This result might be of interest for the foundations of both quantum computation and quantum mechanics.
Landauer also raised the question whether the very concept of measurement necessarily requires an interaction between the system and the apparatus. In the case of the quantum computer, the "apparatus", made of quantum logic gates, interacts with the system, but it is itself part of the system. Quantum computation is then a kind of meta-measurement, in the sense that a quantum computer "measures" itself by its own computational process.
iv) The Truth-Observable is a new logical concept, by which truth judgements coincide with inner reversible measurements. Thus, not only "information is physical" (as Landauer said) but also "inner quantum logic is physical".
v) We showed that only an internal observer could assert the objective reality of a quantum state.
vi) In our view, the holistic cognitive attitude appears as a peculiarity of the internal observer. In fact, the inner quantum truth (the Truth-Observable) is a global truth.
vii) With respect to the "measurement problem" the philosophical approach of this paper stands between Bohr and Einstein's views. In fact, Einstein's realist position relies on the belief that, *under ideal conditions*, measurements behave like *mirrors* [14] in the sense that they reflect an independently existing reality. To us, those *ideal conditions* are realized only in an inner reversible measurement. On the other hand, Bohr's position, which is neo-Kantian relationalism, stands on the assumption that the dependence of properties upon experimental conditions is relational, not causal. To Bohr, it is not meaningless to assign a value to a quantity in a quantum system unless it is measured (this was Heisenberg's position). To us, Bohr's view means that the value of a certain quantity is *reflected* by an imperfect mirror, the imperfection of the resulting image depending on which kind of measurement is performed. In the case of the inner reversible measurement, and only in that case, Einstein and Bohr views coincide, as the mirroring is perfect.

viii) There are some intriguing implications of the Truth-Observable in the Intuitionist, Spinozian and Kantian philosophical thoughts.

Accordingly to intuitionist philosophy, a mathematical truth is such only if it is computable. This is a constructive approach to mathematics and logic. In our case, we can say then that the Truth-Observable is a constructive mathematical concept as it can be computed (or measured, which is the same inside a quantum computer). Spinoza's law of necessity states that the *modes* (humans and the world of things) necessarily follow from *substance* plus attributes. In a modern re-interpretation of Spinoza, R. Zimmermann [15] suggested that substance could be viewed as pre-geometry at the Planck scale. In a quantum-computing universe modelled in terms of non-commutative geometry [5] the internal observer is then a pullback of the modes into the substance. Moreover, the concept of Truth-Observable means that the internal observer finds equality between the physical observable and the logical truth, that is a modern re-interpretation of Spinoza's parallelism between logic and physics [15]. Finally, the internal observer has a contact with a phenomenon (the quantum superposition) which otherwise is considered as a noumenon by our mind, then for him/her, the Kantian noumenon coincides with the phenomenon.

**Acknowledgements**
I wish to thank Pieralberto Marchetti, Giulia Battilotti and Dario Maguolo for useful discussions.